\documentstyle[a4,12pt,epsfig,here]{article}
\begin{document}

\title{\bf Anomalous single production of the fourth SM  
family quarks at Tevatron} 

\author{E. Arik$^{a}$, O. Cak\i r$^{b}$, S. Sultansoy$^{c,d}$ \\
$a$) Bo\u{g}azi\c{c}i University, Faculty of Arts and Sciences, 
Department of Physics,  \\
 80815, Bebek, Istanbul, Turkey \\
$b$) Ankara University, Faculty of Science, Department of Physics, \\
06100 Tando\u gan, Ankara, Turkey \\
$c$) Gazi University, Faculty of Arts and  Sciences, Department of Physics,\\
06500, Teknikokullar, Ankara, Turkey \\
$d$) Azerbaijan Academy of Sciences, Institute of Physics, \\
H. Cavid Av., 33, Baku, Azerbaijan }

\date{ }

\maketitle
\vspace*{2.5cm}

\begin{abstract}
Possible single productions of fourth family $u_4$ and $d_4$ quarks via 
anomalous $q_4qV$ interactions at Tevatron are studied. Signature of such processes are 
discussed and compared with the recent results from Tevatron.
\end{abstract}
\vskip 1.0cm

The flavor democracy enforces the existence of the fourth Standard Model (SM) 
family {\cite{datta, celikel}}. The masses of the fourth family quarks are 
expected to be degenerate and lie between $300$ GeV and $700$ GeV (see {\cite{saleh}} and references therein). \\

It is clear that the pair production of the fourth family quarks will be 
copious at CERN LHC {\cite{prd1,tdr}}. In addition, extra SM generations 
will yield an essential enhancement in Higgs production at LHC and Tevatron 
{\cite{prd2,sn,prd3}}. 
Furthermore, the future lepton colliders will be the best place to observe 
the fourth SM family leptons {\cite{prd4,clic}}.\\

Together with the indirect manifestation of the fourth SM family via 
enhancement of the Higgs boson production, the fourth family quarks can also 
be observed at Tevatron if anomalous $q_4qV$ interactions exist. The 
arguments given in {\cite{frit}} for anomalous interactions of the top quark 
are more valid for $u_4$ and $d_4$ since they are expected to be heavier 
than the top quark.\\ 

In this work, we present a preliminary analysis of the anomalous single 
production of $u_4$ and $d_4$ quarks at Tevatron. The recent observation of 
excess of events with the final state containing $W + 2 , 3$ jets 
{\cite{cdf,apol1,apol2}} can be interpreted as the signature of these 
processes. \\

The effective Lagrangian for the  anomalous interactions between the  
fourth family quarks, ordinary quarks and the gauge bosons $V$ 
($V = \gamma, Z, g$) can be written as follows:

\begin{eqnarray}
L = \frac {\kappa_{\gamma}^{q_i}}{\Lambda} e_q g_e \bar q_4 \sigma_{\mu \nu} (A_{\gamma}^{q_i} + B_{\gamma}^{q_i} \gamma_5) q_i F^{\mu \nu} + 
 \frac {\kappa_Z^{q_i}}{2 \Lambda} g_Z \bar q_4 \sigma_{\mu \nu} 
(A_Z^{q_i}  + B_Z^{q_i}  \gamma_5) q_i Z^{\mu \nu} \nonumber \\
+\,  \frac {\kappa_g^{q_i}}{\Lambda} g_s \bar q_4 \sigma_{\mu \nu} 
(A_g^{q_i}  + B_g^{q_i}  \gamma_5) T^a q_i G^{\mu \nu}_a + h.c. 
\end{eqnarray}  
where $F^{\mu \nu}, Z^{\mu \nu}$, and $G^{\mu \nu}$ are the field strength 
tensors of the photon, $Z$ boson and gluons, respectively; $T^a$ are 
Gell-Mann matrices;
$e_q$ is the charge of the quark; 
$g_e, g_Z$, and $g_s$ are the electroweak, and 
the strong coupling constants respectively.  
$g_Z = g_e/\cos\theta_W \sin\theta_W$ where $\theta_W$ is the Weinberg angle. 
 $A_{\gamma, Z, g}^q$ and 
$B_{\gamma, Z, g}^q$ are the magnitudes of the neutral currents; 
$\kappa_{\gamma, Z, g}$ define the strength of the  anomalous couplings for the neutral currents with a photon, a $Z$ boson 
and a gluon, respectively;  $\Lambda$ is the cutoff scale for the new 
physics. \\  

We assume all the neutral current magnitudes in Eq. (1) to be equal, 
satisfying the constraint $|A|^2 + |B|^2 = 1$ and redefine all anomalous 
couplings:
\begin{equation}
 {\kappa_{\gamma}^{q_i}} =  {\kappa_Z^{q_i}} = {\kappa_g^{q_i}}  = \lambda^{(4-i)} 
\end{equation}
where generation number $i = 1, 2, 3, 4$. We have implemented the new 
interaction vertices into the CALCHEP {\cite{calchep}} package. Two values  
of $\lambda$  are considered ($\lambda = 1, \, 0.5$) as an example. 
Table 1 and 2 (3 and 4) present the branching ratios and the total  decay 
widths of the $u_4 \, (d_4)$ quark via anomalous interactions. Note that the 
branching ratios are independent of $\Lambda$ whereas total decay width 
is proportional to $\Lambda^{-2}$.   
SM decay modes are negligible for 
$\lambda / \Lambda > 0.01/$TeV  due to the small magnitude of 
the extended Cabibbo-Kobayashi-Maskawa matrix 
elements $V_{u_4b}$ and  $V_{d_4t}$ {\cite{ayla}}.  \\

\begin{table}[H]
\begin{center}
\caption{Branching ratios ($\%$) and total decay widths for $u_4$ 
($\lambda = 1 , \Lambda = 1$ TeV).}
\vskip 0.5 cm
\begin{tabular}{|c|c|c|c|c|c|c|c|}
\hline
Mass (GeV)& $gu(c)$ & $gt$ & $Zu(c)$ & $Zt$ & $\gamma u(c)$ &  $\gamma t$ & $\Gamma$ (GeV) \\
\hline
200 & 47 & 0.6 & 2.2 & - & 0.99 & 0.031 & 1.39\\
\hline
250 & 44 & 5.8 & 2.4 & - & 0.93 & 0.12 & 2.91 \\
\hline
300 & 41 & 12 & 2.4 & 0.46 & 0.86 & 0.25 & 5.41 \\ 
\hline
400 & 36 & 19 & 2.3 & 1.1 & 0.77 & 0.41 & 14.26 \\
\hline
500 & 34 & 23 & 2.2 & 1.5 & 0.73 & 0.49 & 29.53 \\
\hline
600 & 33 & 25 & 2.2 & 1.7 & 0.71 & 0.54 & 52.82 \\
\hline
700 & 33 & 27 & 2.2 & 1.8 & 0.69 & 0.57 & 85.69 \\
\hline
\end{tabular}
\end{center}
\end{table}

\begin{table}[H]
\begin{center}
\caption{The same as Table 1, but for $\lambda = 0.5$.}
\vskip 0.5 cm
\begin{tabular}{|c|c|c|c|c|c|c|c|c|c|c|}
\hline
Mass (GeV) & $gu$ & $gc$ & $gt$ & $Zu$ & $Zc$ & $Zt$ & $\gamma u$ & $\gamma c$  &  $\gamma t$  & $\Gamma$ (GeV) \\
\hline
200 & 18 & 72 & 3.7 & 0.86 & 3.4 & -  & 0.38 & 1.5 & 0.079 & 0.06\\
\hline
250 & 13 & 53 & 28 & 0.73 & 2.9 & -  & 0.28 & 1.1 & 0.6 & 0.15 \\
\hline
300 & 9.7 & 39 & 45 & 0.58 & 1.3 & 1.7 & 0.21 & 0.83 & 0.95 & 0.35 \\ 
\hline
400 & 6.9 & 27 & 58 & 0.43 & 1.7 & 3.4 & 0.15 & 0.58 & 1.2 & 1.18 \\
\hline
500 & 5.8 & 23 & 63 & 0.38 & 1.5 & 4.0 & 0.12 & 0.49 & 1.3 & 2.72 \\
\hline
600 & 5.3 & 21 & 69 & 0.35 & 1.4 & 4.3 & 0.11 & 0.45 & 1.4 & 5.15 \\
\hline
700 & 5.1 & 20 & 67 & 0.34 & 1.4 & 4.4 & 0.11 & 0.43 & 1.4  & 8.62 \\
\hline
\end{tabular}
\end{center}
\end{table}

\begin{table}[H]
\begin{center}
\caption{Branching ratios ($\%$) and decay widths for $d_4$ 
($\lambda = 1  , \Lambda = 1$ TeV).}
\vskip 0.5 cm
\begin{tabular}{|c|c|c|c|c|}
\hline
Mass (GeV)& $gd(s,b)$  & $Zd(s,b)$  & $\gamma d(s,b)$  & $\Gamma$ (GeV) \\
\hline
200 & 32 & 1.5 & 0.17 & 2.05\\
\hline
250 & 31 & 1.7 & 0.17 & 4.04 \\
\hline
300 & 31 & 1.9 & 0.17 & 7.01 \\ 
\hline
400 & 31 & 2.0 & 0.17 & 16.68 \\
\hline
500 & 31 & 2.0 & 0.17 & 32.64 \\
\hline
600 & 31 & 2.1 & 0.16 & 56.46  \\
\hline
700 & 31 & 2.1 & 0.16 & 89.71 \\
\hline
\end{tabular}
\end{center}
\end{table}

\begin{table}[H]
\begin{center}
\caption{Same as Table 2 but for $\lambda = 0.5$.}
\vskip 0.5 cm
\begin{tabular}{|c|c|c|c|c|c|c|c|c|c|c|}
\hline
Mass (GeV) & $gd$ & $gs$ & $gb$ & $Zd$ & $Zs$ & $Zb$ & $\gamma d$ & $\gamma s$  &  $\gamma b$  & $\Gamma$ (GeV) \\
\hline
200 & 4.5 & 18 & 72 & 0.22 & 0.86 & 3.4 & 0.024 & 0.096 & 0.38 & 0.22\\
\hline
250 & 4.5 & 18 & 72 & 0.25 & 0.99 & 4.0 & 0.024 & 0.095 & 0.38 & 0.44 \\
\hline
300 & 4.5 & 18 & 72 & 0.26 & 1.1 & 4.2 & 0.024 & 0.095 & 0.38 & 0.76  \\ 
\hline
400 & 4.5 & 18 & 71 & 0.28 & 1.1 & 4.5 & 0.024 & 0.095 & 0.38 & 1.82 \\
\hline
500 & 4.4 & 18 & 71 & 0.29 & 1.2 & 4.6 & 0.024 & 0.094 & 0.38 & 3.57 \\
\hline
600 & 4.4 & 18 & 71 & 0.29 & 1.2 & 4.7 & 0.024 & 0.094 & 0.38 & 6.17 \\
\hline
700 &  4.4 & 18 & 71 & 0.3 & 1.2 & 4.8 & 0.024 & 0.094 & 0.38 & 9.81 \\
\hline
\end{tabular}
\end{center}
\end{table}

Anomalous single production cross sections for $u_4$ and $d_4$ quarks at 
Tevatron are plotted in Figure 1. The upper curves are obtained with  
$\lambda = 1, \Lambda = 2$ TeV and the lower ones with  
$\lambda = 0.5, \Lambda = 1$ TeV. The excess of events with a $W$ and a 
superjet in $W + 2$ jet sample, observed by the CDF collaboration at 
Tevatron, can be explained 
as anomalous single production of $u_4$ with subsequent decay chain 
$u_4 \rightarrow t \, g \rightarrow W  b \,  g$. Let us estimate the number 
of events expected in both senarios mentioned above. 
In the case of $\lambda = 0.5$, $\sigma(p \bar p \rightarrow u_4 X) \times 
BR(u_4  \rightarrow t g) = 0.31$ pb for $m_{u_4} = 400$ GeV. Assuming $40 \%$ 
detector efficiency and $BR(W \rightarrow e \nu + \mu \nu) \approx 0.21$, 
one obtains $2 - 3$ such events for $106$ pb$^{-1}$ integrated luminosity. 
In addition, the same number of events are  expected  from 
$p \bar p \rightarrow \bar u_4 X$.  CDF collaboration has observed $8$ 
events while the SM prediction is $ 2.69 \pm 0.41$ {\cite{cdf}}. 
In the case of 
$\lambda = 1$, one needs to set $\Lambda = 4$ TeV in order to obtain 
similar number of events (for $\lambda = 1\,,\,\sigma \sim \Lambda^{-4}$).\\

The other main decay modes, namely $g u $ and $g c$ 
will give rise to dijet final states. For $m_{u_4} = 400$ GeV,  
$\sigma \times BR(u_4  \rightarrow jj) \approx 0.18$ pb  with 
$\lambda = 0.5$ and $\Lambda = 1$ TeV and $\sigma \times 
BR(u_4  \rightarrow jj) \approx 1.1$ pb with 
$\lambda = 1$ and $\Lambda = 4$ TeV. The CDF dijet analysis {\cite{dijet}} 
lead to an upper limit of $16$ pb at this mass. The contribution of $d_4$ 
to the dijet events is estimated to be $0.6$ pb with 
$\lambda = 0.5$ and $\Lambda = 1$ TeV and $0.75$ pb with 
$\lambda = 1$ and $\Lambda = 4$ TeV. \\

Finally, if we allow the SM decay modes of $d_4$  to be comparable 
to anomalous ones, the superjet events in $W + 3$ jet sample can be explained 
through the following decay chain:  
$d_4 \rightarrow W t \rightarrow W W b $, where one $W$ decays 
leptonically and the other decays into two jets. CDF collaboration has 
observed $5$ events while the SM prediction is $ 1.71 \pm 0.40$ {\cite{cdf}}.\\

In conclusion, the upgraded Tevatron reaching integrated luminosity 
$15$ fb$^{-1}$ can observe the fourth SM family quarks before the LHC, 
provided that the anomalous single production is dominant.

\begin{figure}[H]
\begin{center}
\epsfig{file=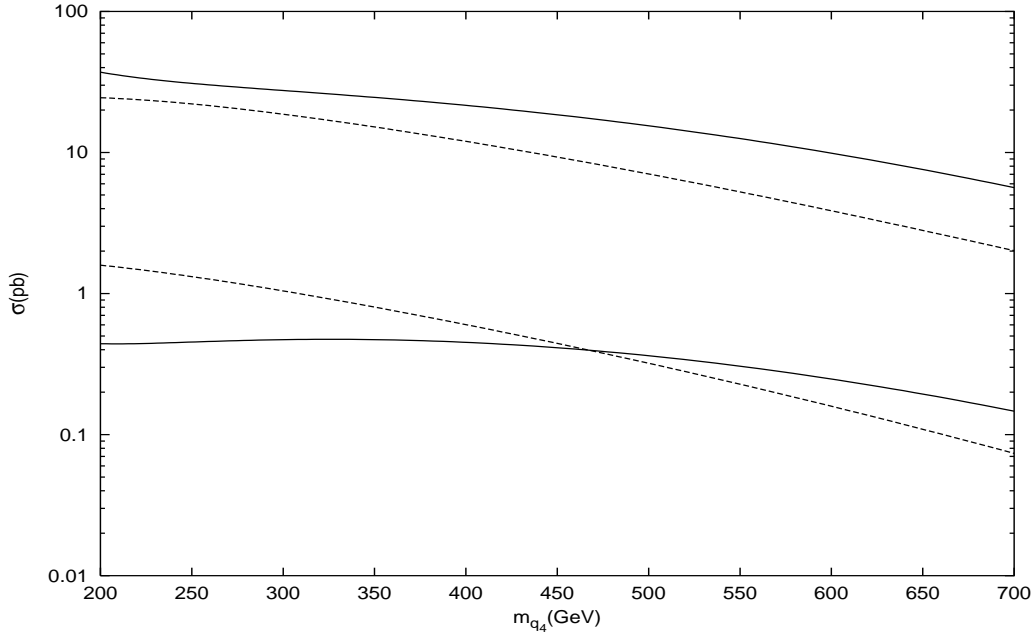,height=8.5cm,width=14cm}
\hfil
\caption{Anomalous single production cross section for $u_{4}$ (solid curves), 
 $d_{4}$ (dashed curves) quarks. Upper (lower) curves correspond to $\lambda = 1,\, \Lambda = 2$ TeV  ($\lambda = 0.5, \, \Lambda = 1$ TeV). }
\label{fig.1}
\end{center}
\end{figure}

\noindent
\begin{center}{\bf Acknowledgments} \\
\end{center}

This work is partially supported by Turkish State Planning Organization 
under the Grant No 2002K120250.\\

\end{document}